\documentclass{article}
\usepackage[utf8]{inputenc}
\usepackage{amsmath,amssymb}
\usepackage{graphicx}
\usepackage[a4paper,top=3cm,bottom=2cm,left=3cm,right=3cm,marginparwidth=1.75cm]{geometry}

\title{On the relation between the astrophysical neutrino fluxes \\ and the cosmic ray fluxes}
\author{ Esteban Roulet\\
Centro At\'omico Bariloche, Comisi\'on Nacional de Energ\'\i a At\'omica\\ 
Consejo Nacional de Investigaciones Cient\'\i ficas y T\'ecnicas (CONICET)\\ 
Av. Bustillo 9500, R8402AGP, Bariloche, Argentina}
\date{}

\begin{document}

\maketitle

\begin{abstract}
Some generalizations of the relation between high-energy astrophysical neutrino and cosmic ray fluxes are obtained, taking into account present results on the cosmic ray spectrum and composition as well as a more realistic modeling of the Galactic and extragalactic cosmic ray components down to PeV energies. It is found that the level of neutrino fluxes measured by IceCube can be consistent with sources that are thin to escaping protons. This could also make it easier for heavier nuclei to be emitted from the sources without suffering excessive disintegration processes.

\end{abstract}

\section{Introduction}

Since high-energy astrophysical neutrinos are expected to be produced mostly through the decay of the charged pions resulting from the interactions of cosmic ray (CR) nuclei with target material at the sources, a relation between the neutrino fluxes and the CR fluxes is expected to hold. The neutrino fluxes should be proportional to the probability for the accelerated CRs to interact at their sources, and hence to the column density of the target material (either photons or gas) that they traverse before exiting. On the other hand, for the CR production to be efficient one expects that, at least at very high energies, the probability for the CRs to escape the sources without being attenuated by interactions be significant, and hence that the sources be {\em thin} to the accelerated CRs. The maximum neutrino flux resulting from thin sources, known as the Waxman-Bahcall (WB) flux limit \cite{wb99,wb01}, is obtained then when the probability for the accelerated CRs to interact at the sources is set to unity, while considering  that the same sources are also responsible for the observed CR flux at the highest energies. This flux level has been used as a guide to estimate the astrophysical neutrino fluxes that could be explored with neutrino observatories, such as IceCube, and remarkably the fluxes that were observed \cite{ic,ic2,ic4} are not far from the WB neutrino flux bound.

The original derivation of the WB flux level was obtained considering that most of the CRs above 10\,EeV (where $1\,{\rm EeV}\equiv 10^{18}\,{\rm eV}$) were protons of extragalactic origin, and the observed fluxes above those energies were used  to normalize the CR production rate.  This rate was then extrapolated to lower energies assuming a constant energy production per logarithmic bin, as results from the $E^{-2}$ spectrum expected from diffusive shock acceleration, under the assumption that below the spectral ankle feature at $\sim 5$\,EeV the CRs were predominantly of Galactic origin \cite{wb99}, or eventually considering that below the ankle  the extragalactic protons contributed at most about 10\% of the observed CR flux \cite{wb01}. 

In this work we reconsider this subject, obtaining a direct relationship between  the differential neutrino fluxes and the differential extragalactic CR fluxes in terms of the modification factor introduced in \cite{be88} to account for the  attenuation of the CR proton fluxes which result from their interaction with the CMB photons. We also discuss the possible contribution from heavier CR nuclei to the neutrino fluxes. This allows us to consider realistic scenarios that account for the present observations relating to the CR fluxes and their composition (i.e. the distribution of their masses).
In particular, the measurements obtained with the Pierre Auger Observatory suggest that the CR composition becomes increasingly heavier above the ankle, and do not support the presence of a significant component  of protons above 5\,EeV \cite{combinedfit}, what is in tension with the assumptions underlying the WB limit. Moreover, a significant fraction of the protons which are inferred to be present in the CR flux at energies of few EeV may find their origin as secondaries from  higher energy CR nuclei that get photo-disintegrated as they travel to us through intergalactic space, and hence would not contribute to the neutrino production at the sources.

On the other hand, there are indications that the knee feature in the CR spectrum, at about 4~PeV, is related to the suppression of the Galactic proton component. In scenarios in which the suppression effects depend on the rigidity of the particles, and hence on the ratio $E/Z$ with $Z$ being the particle charge, the Galactic Fe component should become suppressed above an energy of about 100~PeV, which actually coincides with the observed steepening at the second knee of the spectrum. The presence of a predominant extragalactic component above this energy is then likely \cite{mo19}, and also the lack of significant anisotropies at EeV energies associated with the Galactic center or the Galactic plane disfavors a sizeable contribution of Galactic origin at those energies \cite{anisora}.
We also note that the extragalactic component relevant below a few EeV may arise from a different source population than the  one dominating at higher energies, and hence to use the latter to normalize the flux of the lower energy component may not be justified. Moreover, the relation between the potential neutrino fluxes and the observed CR fluxes could also be affected by the presence  of a magnetic horizon, which could suppress the observed CR spectrum from far away sources for decreasing energies, as we will also discuss.

\section{The production of neutrinos from CR interactions}
CRs that get accelerated at a source, as for instance in the jets of active galactic nuclei (AGNs) or gamma ray bursts (GRBs), may interact with ambient radiation, such as thermal photons or synchrotron photons from co-accelerated electrons, or alternatively with the ambient gas which is present at or around the acceleration region (for reviews see \cite{me17,ah19}). These two possibilities  are usually referred to  as the $p\gamma$ and the $pp$ mechanisms respectively (focusing first on the interactions of CR protons). Through the production of pions and their subsequent decays these processes can give rise to high-energy neutrinos and photons. Although the production of some heavier weakly decaying hadrons can also contribute to the neutrino production, the dominant channel is in general from the charged pion decay chain
\begin{equation}
    \pi^{+}\to \mu^{+}\nu_\mu\ \ \ ,\ \ \ \mu^{+}\to \bar{\nu}_\mu e^{+}\nu_e,
\end{equation}
and similarly for the $\pi^{-}$ charged conjugate channel, while the gamma rays are mostly produced through the neutral pion decay $\pi^0\to\gamma\gamma$, with a subdominant contribution from the isoscalar $\eta$ decays.

In the $p\gamma$ interaction, a very important channel for $\pi$ production is through the $s$-channel $\Delta(1232)$ resonance, $p\gamma\to \Delta\to n\pi^{+}$  (or $p\pi^0$), but there are also contributions from non-resonant direct $\pi^{+}$ exchange, from higher-mass resonances and, at larger center of mass energies, from multi-pion production processes.  While at the $\Delta$-resonance the number of $\pi^0$ produced is about twice that of $\pi^{+}$, the direct channel
produces mostly $\pi^{+}$  while the multi-pion production  leads to comparable numbers of $\pi^{+}$, $\pi^{-}$ and $\pi^0$. The rate of these interactions, and hence the resulting neutrino spectrum, depends not only on the proton energies but also on the spectrum of the target photons, and due to the threshold energy for the pion production it only occurs for $E_p>(2m_pm_\pi+m_\pi^2)/4\varepsilon_\gamma\simeq 70/(\varepsilon_\gamma/{\rm keV})$~TeV. The details of the interactions can be described for instance  using the SOPHIA   package \cite{sophia} or using approximate analytical parametrizations \cite{hu10}. 

In the $pp$ interactions the pion production is  instead possible down to GeV proton energies, and the number of pions that get produced steadily increases with energy, being of order $10^2$ for PeV proton energies. The details of the non-perturbative hadronic interactions  can be described with specific  models, such as Sibyll \cite{sibyll} or EPOS \cite{epos}.  

The fraction of the proton energy  which is transferred to the produced pions is given by the inelasticity $\kappa$ of the interaction, with $E_{\rm pions}\simeq \kappa E_p$. For the $p\gamma$ interaction near the $\Delta$-resonance peak, one has that $\kappa\simeq 0.2$, while at much higher energies one has that $\kappa\simeq 0.5$, a value similar to that  associated to the $pp$ interactions. The pions produced near the $\Delta$-resonance typically carry about 20\% of the proton energy, but one has to keep in mind that  in the regime in which many pions get produced, most of them have energies much smaller than $E_p$, with only the most energetic leading pion  having on average an energy close to $E_p/5$.
If the CR spectrum is very steep (i.e. soft), as in the case of a power-law d$\Phi_{\rm CR}/{\rm d}E\propto E^{-\alpha}$ with $\alpha>2.5$, in the regime of multipion production the lower energy pions play a subdominant role, being buried under the larger flux from leading pions arising from less energetic CRs. Instead for flatter (i.e. harder) CR spectra the large number of lower energy pions do play a significant role. 

In order to obtain the neutrino spectrum resulting from the CR interactions, one has to make the convolution of the pion yield from protons of a given energy with the CR proton spectrum and then compute the neutrino yield from the decaying pions, eventually accounting for energy losses of the charged pions and muons. These losses, which are mostly due to synchrotron radiation, are only relevant at very high energies ($E_\nu>{\rm PeV}$, so that the decaying particles live long enough) and if the magnetic fields in the interaction region are very large, and hence we will ignore them for simplicity. 
Since the pions ultimately decay into four leptons with negligible masses (e.g. $\pi^{+}\to e^{+},\,\nu_e,\,\bar{\nu}_\mu$ and $\nu_\mu$), one has that typically $E_{\nu_i}\simeq E_\pi/4$, but anyhow all the neutrino energies have a continuous distribution in the observer's frame.

In the regime of multipion production, which is especially relevant in $pp$ scenarios, the median energy $\bar{E}_p$ of the protons giving rise to a given neutrino energy turns out to be significantly larger than the  usually adopted naive estimate $E_p\simeq 20E_\nu$ (which would correspond to assuming that $E_\pi\simeq E_p/5$), and it depends on the CR spectral slope and on the neutrino energy considered \cite{ro21}. For instance, in the case of PeV neutrinos arising in $pp$ scenarios, for $\alpha=2$   one has that $\bar{E}_p/E_\nu\simeq 50$ while for  $\alpha=2.3$ one has that $\bar{E}_p/E_\nu\simeq 30$.  Given the long tails in the distribution of the  parent proton  energies, the associated average proton  to neutrino energy ratios are actually even higher than the median ones.
In the $p\gamma$ scenarios, due to the fact that usually the photon target spectrum is steep, the abundant low energy photons make the contribution of the $\Delta$-resonance to stay relevant even for very-high proton energies, and hence the impact of the multipion production is not so pronounced as in the $pp$ scenarios (moreover, even for proton energies a factor of 100 above the $\Delta$-resonance the pion multiplicities associated to interactions with photons are still of only a few). This makes the approximation that $E_p\simeq 20E_\nu$ to hold to a better approximation in $p\gamma$ scenarios. However, one also finds in this case, as a consequence of the threshold for pion production which requires that  ${E}_p/E_\nu > 20/(E_\nu/100\,{\rm TeV})/(\varepsilon_\gamma/30\,{\rm eV})$, that the  low energy tail of the neutrino spectrum may only be produced by protons with associated values of $\bar{E}_p/E_\nu$ significantly exceeding the value  20 \cite{ro21}, specially if the sources have a narrow photon spectra so that the high-energy target photons are very rare.

When estimating  the neutrino spectrum from the underlying proton spectrum giving rise to it, we will hence consider for simplicity that the  neutrinos with energy $E_\nu$ can be directly related to the flux of CR protons with energy  $E_p=\rho E_\nu$, adopting the default value $\rho\simeq 20$ but analysing also the impact of having larger values of $\rho$, what should be particularly relevant in the case of $pp$ scenarios.

Considering then that the CRs are protons and that they all traverse a similar amount of target material at the sources, characterized by a column density $N_{\rm t}=n_{\rm t} L_{\rm int}$, with $L_{\rm int}$ being the distance traversed  in the interaction region of density $n_{\rm t}$, the neutrino emissivity per source  $q_\nu$ (number of neutrinos per unit time) will be related to the proton one $q_p$ through
\begin{equation}
    \sum_iE_{\nu}\frac{{\rm d}q_{\nu_i}}{{\rm d\,ln}E_{\nu}}(E_{\nu})\simeq \frac{3}{4}R_{\rm ch}\left[1-\exp(-N_{\rm t}\sigma_{p{\rm t}})\right]\kappa E_p\frac{{\rm d}q_{p}}{{\rm d\,ln}E_p}(E_p=\rho E_\nu),
    \label{qnuvsqp}
\end{equation}
where $R_{\rm ch}\equiv (M_{\pi^{+}}+M_{\pi^{-}})/(M_{\pi^{+}}+M_{\pi^{0}}+M_{\pi^{-}})$ is the fraction of the pions that are charged, with $M_\pi$ being the corresponding multiplicities, and the factor 3/4 takes into account that approximately $E_\pi/4$ gets transferred to electrons. One  denotes the fraction of the energy of the  protons which is transferred to the pions as $F_\pi\simeq \kappa P$, with $P\equiv \left[1-\exp(-N_{\rm t}\sigma_{p{\rm t}})\right]$ being the probability for the proton to interact in the source environment. An upper limit for the neutrino flux from a thin source is usually obtained setting $F_\pi=1$, as considered originally by Waxman and Bahcall, with the most likely neutrino flux being actually a factor $F_\pi$ smaller than this upper bound \cite{wb99,wb01}.

Denoting the emissivities per unit volume as $Q=n_{\rm s}q$, with $n_{\rm s}$ being the CR source density, and considering a possible cosmological evolution with redshift parametrized as
\begin{equation}
    n_{\rm s}(z)=f(z)n_{\rm s}(0),
\end{equation}
one can compute the present neutrino density integrating the contribution from sources at different redshifts, from the present time ($z=0$) up to a maximum value $z_{\rm m}$, as
\begin{equation}
    \frac{{\rm d}n_\nu}{{\rm d}E_\nu}(E_\nu)=\int_0^{z_{\rm m}}{\rm d}z\,\left| \frac{{\rm d}t}{{\rm d}z}\right|\frac{{\rm d}Q_\nu}{{\rm d}E'_\nu}\frac{{\rm d}E'_\nu}{{\rm d}E_\nu},
\end{equation}
where, due to the cosmological expansion, one has that $E'_\nu=(1+z)E_\nu$, and
\begin{equation}
    \left| \frac{{\rm d}t}{{\rm d}z}\right|=\frac{1}{H_0(1+z)\sqrt{(1+z)^3\Omega_m+\Omega_\Lambda}},
\end{equation}
in terms of the present day Hubble constant $H_0\simeq 70$~km\,s$^{-1}$Mpc$^{-1}$, with $\Omega_m\simeq 0.3$ and $\Omega_\Lambda\simeq 0.7$ being the matter and vacuum energy contributions to the critical density at present. 

Combining the above expressions with eq.\,(\ref{qnuvsqp}), one gets that 
\begin{eqnarray}
    \sum_i\frac{{\rm d}n_{\nu_i}}{{\rm d}E_{\nu}}(E_{\nu})&=&\sum_i\int_0^{z_{\rm m}}{\rm d}z\,(1+z)\left| \frac{{\rm d}t}{{\rm d}z}\right|\frac{{\rm d}Q_{\nu_i}}{{\rm d}E'_\nu}(z,E'_\nu)\nonumber\\
    &\simeq& \frac{3}{4}R_{\rm ch}F_\pi \int_0^{z_{\rm m}}{\rm d}z\,(1+z)\left| \frac{{\rm d}t}{{\rm d}z}\right|\frac{{E'}_p^2}{{E'}_\nu^2}\frac{{\rm d}Q_{p}}{{\rm d}E'_p}(z,E'_p=\rho E'_\nu)\\
    &\simeq &\frac{3}{4}R_{\rm ch}F_\pi\rho^2 \int_0^{z_{\rm m}}{\rm d}z\,(1+z)\left| \frac{{\rm d}t}{{\rm d}z}\right|f(z)\frac{{\rm d}Q_{p}}{{\rm d}E_p}(z=0,E_p=\rho(1+z) E_\nu).\nonumber
    \label{nnuvsqp}
\end{eqnarray}
In particular, focusing on the case of a power-law spectrum such that d$Q_p/{\rm d}E_p\sim E_p^{-\alpha}$, one gets
\begin{equation}
     E_\nu^2 \sum_i\frac{{\rm d}n_{\nu_i}}{{\rm d}E_{\nu}}(E_{\nu})\simeq \frac{3}{4}R_{\rm ch}F_\pi\left[ E_p^2\frac{{\rm d}Q_{p}}{{\rm d}E_p}(z=0,E_p)\right]_{E_p=\rho E_\nu} \int_0^{z_{\rm m}}{\rm d}z\,(1+z)^{1-\alpha}\left| \frac{{\rm d}t}{{\rm d}z}\right|f(z).
\end{equation}  
One can introduce the evolution factor
\begin{equation}
    \xi_z\equiv \frac{1}{t_{\rm U}}\int_0^{z_{\rm m}}{\rm d}z\,(1+z)^{1-\alpha}\left| \frac{{\rm d}t}{{\rm d}z}\right|f(z),
\end{equation}
with $t_{\rm U}=\int_0^\infty {\rm d}z\,|{\rm d}t/{\rm d}z|\simeq 13.5$~Gyr being the lifetime of the universe, to get 
\begin{equation}
     E_\nu^2 \sum_i\frac{{\rm d}n_{\nu_i}}{{\rm d}E_{\nu}}(E_{\nu})\simeq \xi_z\frac{3}{4}R_{\rm ch}F_\pi\left[ E_p^2\frac{{\rm d}Q_{p}}{{\rm d}E_p}(z=0,E_p)\right]_{E_p=\rho E_\nu} t_{\rm U},
\end{equation} 
which generalizes the expression obtained for the case $\alpha=2$ in \cite{wb99}. For $\alpha\simeq 2$ the evolution factor is $\xi_z\simeq 0.5$ in the absence of evolution (i.e. for $f(z)=1$) and it is about 2.5 for the case of a strong source evolution similar to that of the star formation rate, for which we adopt $f(z)\propto (1+z)^{3.4}$ for $z<0.98$,  $f(z)\propto (1+z)^{-0.26}$ for larger $z$ up to $z=4.48$ and then a much steeper decay \cite{ho06}.

The present day CR emissivity can be related, after  accounting for the attenuation effects due to the interactions with the CMB photons and the redshift losses,  to the observed CR spectrum at ultrahigh energies. In particular, ref.~\cite{ka09} obtained, under the assumption that all CRs are protons,  the estimate
\begin{equation}
     \left[ E_p^2\frac{{\rm d}Q_{p}}{{\rm d}E_p}(z=0,E_p)\right]_{E_p=30\,{\rm EeV}}\simeq  \frac{1}{t_{\rm eff}}\left[ E_p^2\frac{{\rm d}n_{\rm CR}}{{\rm d}E_p}(E_p)\right]_{E_p=30\,{\rm EeV}}\simeq 4\times 10^{43}\,{\rm \frac{erg}{Mpc^3yr}},
\end{equation} 
with the effective attenuation time at 30\,EeV (in which case pair production dominates the energy losses)  being $t_{\rm eff}\simeq 2.5$~Gyr for $\alpha=2$. Given the measured value of the CR density at this energy of $E^2{\rm d}n_{\rm CR}/{\rm d}E\simeq 3\times 10^{-21}$~erg\,cm$^{-3}$, one gets a CR injection rate at present of $E^2{\rm d}Q_{\rm CR}/{\rm d}E(z=0,E=30\,{\rm EeV})\simeq 4\times 10^{43}$~erg\,Mpc$^{-3}$yr$^{-1}$. Note that since at ultrahigh energies only the local universe contributes to the observed  CR flux, the cosmological source evolution has no significant  impact on this value, although the result can depend in principle on the CR spectral shape, since due to the energy losses the CRs observed at a given energy originated at higher ones.

The original WB bound was obtained considering that the CR flux above 10\,EeV consisted predominantly of extragalactic protons having a source spectrum d$q_p/{\rm d}E_p\propto E_p^{-2}$, and having in mind that the CR flux below the ankle was mostly due to a Galactic component which was not contributing to the neutrino fluxes. Normalizing the extragalactic flux to the energy density in CRs determined at 30\,EeV, and extrapolating this flux down to lower energies, the WB flux  limit reads
\begin{equation}
     E_\nu^2 \sum_i\frac{{\rm d}\Phi_{\nu_i}}{{\rm d}E_{\nu}}(E_{\nu})\simeq  \xi_z\frac{R_{\rm ch}}{2/3}F_\pi\,1.3\times 10^{-8}\, \frac{\rm GeV}{\rm cm^2\,s\,sr},
\end{equation}
where $\Phi=(c/4\pi)n$. This value is consistent with what obtained in \cite{ka09}.

We want here also to point out that, under the assumption that the extragalactic CR flux is dominated by protons, there is a direct relationship between the expected astrophysical neutrino fluxes and the extragalactic component of the observed CR spectrum,
and this  relation depends on the attenuation factor $\eta$ introduced in ref.~\cite{be88}. To see this, consider the expression for the CR proton density
\begin{equation}
    \frac{{\rm d}n_p}{{\rm d}E}(E_p)=\int_0^{z_{\rm m}}{\rm d}z\,\left| \frac{{\rm d}t}{{\rm d}z}\right|\frac{{\rm d}Q_p}{{\rm d}E_{\rm g}}\frac{{\rm d}E_{\rm g}}{{\rm d}E_p},
\end{equation}
where $E_g$ is the original energy with which the CR exited the source in order to arrive to Earth with energy $E_p$. The attenuation factor $\eta(E_p)$ is defined  as the ratio between the actual CR spectrum and the one that would have been obtained from the same sources but in the absence of interactions during the propagation, just accounting for redshift effects, i.e.
\begin{equation}
    \eta(E)\equiv \frac{{\rm d}n_p/{\rm d}E(E)}{{\rm d}n_p^{\rm NI}/{\rm d}E(E)},
\end{equation}
where 
\begin{equation}
    \frac{{\rm d}n_p^{\rm NI}}{{\rm d}E}(E)=\int_0^{z_{\rm m}}{\rm d}z\,\left| \frac{{\rm d}t}{{\rm d}z}\right|\frac{{\rm d}Q_p}{{\rm d}E'}\frac{{\rm d}E'}{{\rm d}E},
    \label{dnpnide}
\end{equation}
with $E'=(1+z)E$. Noting that the integral in eq.\,(\ref{dnpnide}) is similar to that in eq.\,(\ref{nnuvsqp}), we then obtain that
\begin{equation}
     \sum_i\frac{{\rm d}n_{\nu_i}}{{\rm d}E_{\nu}}(E_{\nu})\simeq \frac{3}{4}R_{\rm ch}F_\pi\rho^2\frac{{\rm d}n_{p}}{{\rm d}E_p}(\rho E_\nu)\frac{1}{\eta(\rho E_\nu)}.
     \label{newbound}
\end{equation}
This expression provides a link between the expected neutrino spectrum and the extragalactic CR proton spectrum, valid for different energies, without the need of introducing separately $\xi_z$ and $t_{\rm eff}$ and to rely on particular assumptions about the energy dependence for the injection rate. Moreover, the attenuation factor $\eta$ has been found to be quite insensitive to the assumed CR spectral index, depending mostly on the source evolution adopted, and analytical expressions for it as a function of the energy have been obtained for different source evolution scenarios \cite{mo20}. 

\begin{figure}[t]
\centering
\includegraphics[scale=1.3,angle=0]{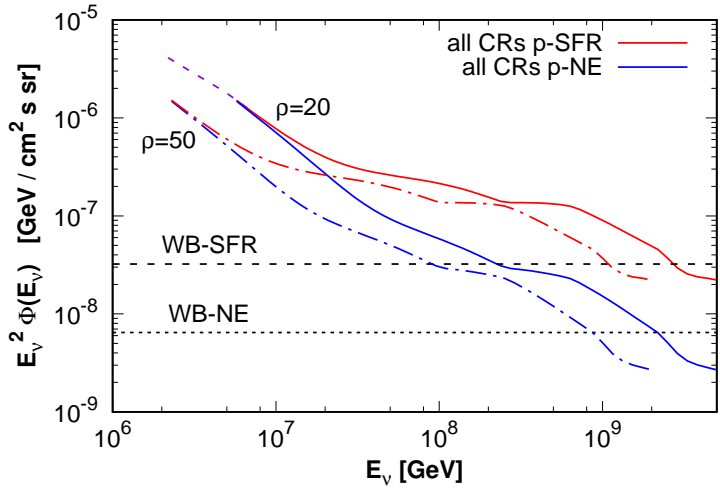}
\caption{upper bounds on the total astrophysical neutrino flux from thin sources for sources with no evolution (NE) or evolving as the SFR. Horizontal lines are for the WB bound, while the other lines assume that all CRs down to the second-knee are protons for $\rho=20$ (solid) or 50 (dot-dashed).}
\label{fwb}
\end{figure}

In Figure~\ref{fwb} we show the astrophysical neutrino flux upper bound obtained considering $F_\pi=1$ (and for $R_{\rm ch}=2/3$), assuming that the CR sources just emit protons and under different hypothesis. The dashed horizontal lines correspond to the WB bound  for an $E^{-2}$ proton spectrum at the sources  normalized to the flux observed at 30\,EeV and adopting the  cosmological source evolution following the star formation rate, with $\xi_z=2.5$ (SFR, long dashes), or for the non-evolving sources with $\xi_z=0.5$ (NE, short dashes). The other lines show the bounds resulting from eq.\,(\ref{newbound})  considering that all CRs observed down to the second-knee are produced in extragalactic proton sources, for the scenario with no source evolution   (red lines) and for the one following the star formation rate (blue lines). The solid lines correspond to adopting $\rho=20$, while the
dot-dashed lines are the results obtained adopting $\rho=50$ (while the WB bound is independent of the value adopted). 
For the CR flux we used the fit recently obtained for energies above 2\,EeV in \cite{augerspec}, extending it as a power-law with index 3.4 down to the second-knee at about 0.1\,EeV. This figure illustrates then the potential impact of the underlying assumptions regarding the extragalactic CR sources and their cosmological evolution. The bounds obtained assuming that all CRs down to the second knee are extragalactic protons  represent some kind of maximum upper bound on the astrophysical neutrino fluxes from thin sources, since part of the CR flux  may be produced in Galactic sources. Moreover, the assumption that the CR fluxes are entirely due to protons and that the proton composition dominates up to the highest energies are not supported by observations. We hence turn now to discuss the case of heavier primaries to see how their presence could affect the results.

\section{The impact of a heavy CR composition}

When nuclei interact with a target, either gas or photons, the neutrino production involves the production of pions, and hence in the nucleus rest frame the target ($p$ or $\gamma$) should transfer an energy larger than the pion mass, and this implies that the interaction actually  takes place at the level of the individual nucleons.  The scattered hadrons subsequently initiate an intra-nuclear cascade and eventually exit the nucleus after depositing some energy that heats the nucleus. This leads to an additional emission of nucleons  and/or nuclear fragments through an evaporation process, contributing to the spallation or photo-disintegration respectively of the initial nucleus.\footnote{For nuclei heavier that Fe, which are not relevant for the present discussion, also fission processes may take place.}

In the case of interactions with photons, the nuclear disintegration can also take place for photon energies up to an order of magnitude below the threshold for pion production, through the excitation of the giant dipole resonance  in which the protons in the nucleus oscillate collectively against the neutrons. In particular, if the photon spectrum is steep so that the low energy photons are very abundant, this channel, having a cross section comparable to that of pion production,  could give rise to a very strong disintegration of the CR heavy nuclear component in sources where the optical depth for pion production is large ($F_\pi$ close to unity). As a consequence, in this kind of scenarios the requirement to have a neutrino production close to the upper bound for thin sources may not be compatible with producing at the same time a significant amount of heavy CRs.\footnote{In particular, requiring that nuclei not be photodisintegrated at all energies and adopting an $E^{-2}$ spectrum normalized at UHE would lead to a neutrino flux upper bound even smaller than the WB flux \cite{mu10}.}

Regarding the cross section for neutrino production off nuclei with mass number $A$, 
the simplest approach is to consider the superposition model, in which one assumes that the target $t$ can interact with each nucleon  independently, so that $\sigma_{tA}(E)\simeq A\sigma_{tp}(E/A)$ (keeping in mind that the pion production happens at similar rates for interactions with protons or neutrons). However, given that there are final state interactions and nuclear shadowing effects, a more realistic approach is to consider that 
\begin{equation}
\sigma_{tA}(E)\simeq A^a\sigma_{tp}(E/A),
\end{equation}
where $a$ is a parameter which is expected to be between 2/3 and 1, with the value of 2/3 corresponding to an interaction scaling as the geometrical cross sectional area of the nucleus while $a=1$ corresponds to the superposition model.
The exponent $a$ may actually be energy dependent, shifting for instance  in the case of target photons  from 1 to 2/3 as the energy increases \cite{mo19b}, and some authors (see e.g. \cite{le63,jo14}) have adopted the intermediate value $a=3/4$ when considering the interactions with target protons. 

The neutrino production is hence obtained using the above mentioned scaling for the total cross section and then computing the neutrino yield from the interactions with nucleons with energy $E/A$. For instance, for the interactions with a H target\footnote{If heavier nuclei such as He were also present, a good approximation is to consider an effective H target density $N_t\simeq N_{\rm H}(1+4N_{\rm He}/N_{\rm H})$.}, and considering for simplicity an energy independent value of the exponent $a$ and that all nuclei traverse the same target column density $N_t$, one has (in the 
limit of thin sources)
\begin{equation}
       \frac{{\rm d}q_{\nu_i}}{{\rm d}E_\nu}(E_\nu)=N_t\sum_AA^a\int_{AE_\nu}^\infty {\rm d}E\sigma_{pp}(E/A)\frac{1}{E/A}F_{\nu_i}(x_\nu,E/A)\frac{{\rm d}q_A}{{\rm d}E}(E),
    \label{dnude}
\end{equation}
with the functions $F_{\nu_i}(x=E_\nu/E_p,E_p)/E_p$
describing the yield of neutrinos carrying a fraction $x=E_\nu/E_p$ of the energy $E_p$ of an incident proton, with parameterizations of the functions $F_{\nu_i}$ being available for instance in \cite{ke06}. Analogous expressions can be written for the interactions with photons, with the results depending in this case on the spectrum of the target photons besides than on the CR spectrum \cite{ke08}.  Note that the expression above assumes that the neutrino energy distribution  resulting from the interaction with the nucleons inside a nucleus is the same as that
for interactions with H nuclei, just the overall normalization is reduced by the factor $A^{a-1}$ that accounts for the nuclear shadowing effects, i.e. for the possible departure of the total nuclear cross section from the simple superposition model.

The total CR emissivity in the source is
\begin{equation}
    \frac{{\rm d}q_{\rm CR}}{{\rm d}E}(E)=\sum_A \frac{{\rm d}q_A}{{\rm d}E}(E),
\end{equation}
with
\begin{equation}
    \frac{{\rm d}q_A}{{\rm d}E}(E)=f_A(E) \frac{{\rm d}q_{\rm CR}}{{\rm d}E}(E)
\end{equation}
in terms of the fractional contribution $f_A(E)$ of nuclei with mass number $A$ to the total CR flux at energy $E$.  One then finds, in scenarios in which the targets are protons,  that
\begin{equation}
       \frac{{\rm d}q_{\nu_i}}{{\rm d}E_\nu}(E_\nu)=N_t\sum_AA^{a+1}\int_{AE_\nu}^\infty \frac{{\rm d}E}{E}\sigma_{pp}(E/A)F_{\nu_i}(AE_\nu/E,E/A)f_A(E)\frac{{\rm d}q_{\rm CR}}{{\rm d}E}(E).
    \label{dnude2}
\end{equation}

In the case of a power-law CR spectrum with d$q_A/{\rm d}E\sim E^{-\alpha}$, in which case also the fractions $f_A$ are independent of energy, one will have that
\begin{equation}
       \frac{{\rm d}q_{\nu_i}}{{\rm d}E_\nu}(E_\nu)=N_t\sum_Af_AA^{a+1-\alpha}\int_{E_\nu}^\infty \frac{{\rm d}E_n}{E_n}\sigma_{pp}(E_n)F_{\nu_i}(E_\nu/E_n,E_n)\frac{{\rm d}q_{\rm CR}}{{\rm d}E}(E_n),
    \label{dnude3}
\end{equation}
where $E_n\equiv E/A$ is the energy per nucleon. This means that the neutrino emissivity, d$q_\nu/{\rm d}E_\nu$, is related to the one that would be obtained under the assumption that all CRs are protons,  d$q_\nu^{({\rm CR}p)}/{\rm d}E_\nu$, by
\begin{equation}
    \frac{{\rm d}q_\nu}{{\rm d}E_\nu}\simeq  \frac{{\rm d}q_\nu^{({\rm CR}p)}}{{\rm d}E_\nu}\sum_Af_AA^{a+1-\alpha}.
\end{equation}
Let us note that this relation will also hold in the case of target photons.
If we adopt for instance the superposition model, corresponding to $a=1$, we see that if the spectral index is $\alpha=2$ the neutrino spectrum will coincide with that obtained assuming that all CRs are protons, but otherwise some changes are expected, and in particular for steeper spectra ($\alpha>2$) one expects a smaller resulting neutrino flux than under the proton only assumption. This can be understood because what is relevant to produce a neutrino of energy $E_\nu$ is the density of CR protons with energies about $\rho E_\nu$, while for nuclei it is their densities at energies $A\rho E_\nu$, which could be very suppressed for large $A$ if the spectrum is steep.

An alternative way to find an expression for the neutrino flux, in the spirit of eq.\,(\ref{newbound}), is to compute the density of nucleons $n_n$ in the CR flux using that
\begin{equation}
    \frac{{\rm d}n_n}{{\rm d\,ln}\,E}(E)=\sum_A A \frac{{\rm d}n_A}{{\rm d\,ln}\,E}(AE),
\end{equation}
so that
\begin{equation}
    \frac{{\rm d}n_n}{{\rm d}E}(E)=\sum_A A^2 \frac{{\rm d}n_A}{{\rm d}E}(AE).
\end{equation}
One can then compute the resulting neutrino flux summing up the contributions from the interactions with the nucleons weighted by the factor $A^{a-1}$ for each nuclear component in order to account for the corresponding scaling of the associated cross section. This leads  to 
\begin{equation}
     \sum_i\frac{{\rm d}n_{\nu_i}}{{\rm d}E_{\nu}}(E_{\nu})\simeq \frac{3}{4}R_{\rm ch}F_\pi\rho^2\sum_AA^{a+1}\frac{{\rm d}n_A}{{\rm d}E}(\rho AE_\nu).
     \label{newbounda}
\end{equation}
 This expression holds for the local densities of neutrinos and CRs as long as one can ignore the CR attenuation effects during the propagation (just keeping the adiabatic losses), so that d$n_A/{\rm d}E$ is the flux that would be observed in the absence of interactions. This will be justified because we  will be dealing with the inferred power-law fluxes normalized at  energies low enough so that the interactions are not relevant. Attenuation effects on the CRs can then be accounted for by multiplying   their fluxes by the corresponding high-energy suppression factors.

 To illustrate the implications of the presence of a heavy component in the CR flux, we discuss now how the astrophysical neutrino flux upper bounds would be modified if one adopts a scenario for the high-energy CRs which is in agreement with the present determinations of their spectrum and composition. We consider the scenario discussed in ref.~\cite{mo20}, which involves two different CR extragalactic source populations that have mixed compositions. The first component dominates the spectrum at energies above few EeV, while the second one, having a steeper spectrum, dominates it for lower energies and down to about 0.1\,EeV, with the Galactic component becoming dominant below that energy. Each extragalactic component involves a power-law spectrum at the sources with a rigidity dependent exponential cutoff. The observed  fluxes are normalized at low energies, for which the attenuation is negligible, so that the nuclear fractions considered will be the same as those present in the sources.

This kind of mixed composition scenario also allows us to discuss another relevant ingredient, which is the fact that the photo-disintegration of the heavier nuclei, that takes place as they travel from their sources through the CMB and EBL radiations, can lead to the production of a significant amount of secondary protons, which may even dominate the observed spectrum at energies of few EeV. These secondary protons however do not contribute to the astrophysical neutrino production, and the cosmogenic neutrinos they may eventually produce as they travel to us  and eventually interact with background photons are expected to be negligible. Hence,  the estimation of the bounds on the astrophysical neutrino fluxes will be based on the inferred  fluxes of primary nuclei  rather than  being directly related to the observed CR fluxes.

In addition, in this scenario the presence of intergalactic turbulent magnetic fields, combined with the finite number density of the CR sources, leads to a magnetic horizon effect that suppresses the extragalactic flux reaching the Earth at low energies. This effect appears when the travel time for the CRs diffusing through the magnetic fields becomes, even for the closest sources, larger than the age of the universe (or the lifetime of the sources) \cite{le05,be07,gl08}. This is particularly relevant for the component dominating at the highest energies, arising from more extreme (and hence less abundant) sources, for which this effect can make the observed spectrum for each mass component to become very hard at rigidities below few EV,  as seems to be required by observations, even if their actual  spectrum at the sources is softer \cite{difu1}. A crucial point is that the neutrino production depends on the inferred CR source spectrum, which in this case can qualitatively differ from the observed one, and can hence be significantly enhanced at the energies where the magnetic suppression is strong.

Note also that the extension of the extragalactic CR source flux below 0.1\,EeV is quite uncertain, since those CRs may not reach us due to the magnetic horizon effect. The usual expectation from the Fermi diffuse shock acceleration mechanism that the source spectral index should be close to $\alpha=2$ could be used as a guide, with the steeper effective spectral index $\alpha\simeq 3.4$ which is required above 0.1\,EeV and up to the ankle eventually resulting from the superposition of many sources with harder spectrum that have a distribution of cutoff energies \cite{ka06}. We will then consider two possible extrapolations for the spectrum of the extragalactic sources below 0.1\,EeV and down to PeV energies (and hence for neutrino energies down to about 50~TeV), that should encompass a wide range of possible source scenarios. In the first we consider that the spectral index below 0.1\,EeV of the low-energy extragalactic component, $\alpha_0$, keeps the same value 3.4 as at higher energies, and we also consider another case with $\alpha_0=2$. This last case could represent a scenario in which the extragalactic sources have  $\alpha\simeq 2$, with most of them being able to reach maximum energies larger than about 0.1\,EeV, but having then a distribution of cutoff energies above that threshold. We will actually consider in this case a rigidity dependent break for the different source components at energies $0.1Z$\,EeV.

\begin{figure}[t]
\centering
\includegraphics[scale=1.3,angle=0]{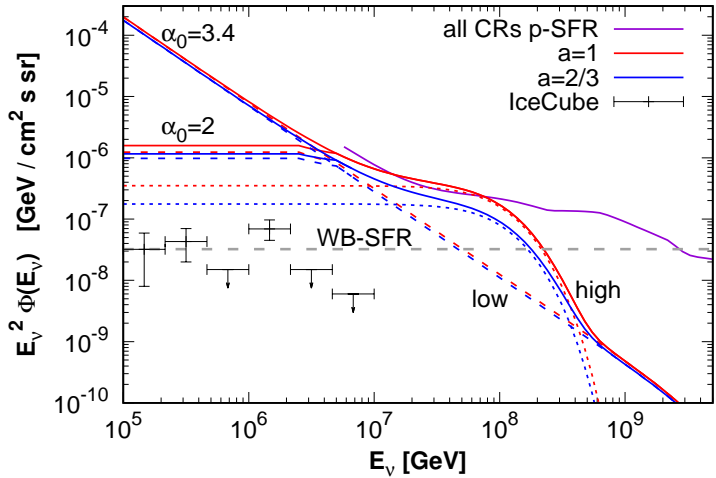}
\caption{upper bounds on the total astrophysical neutrino flux from thin sources in the SFR-NE scenario of \cite{mo20} under different assumptions and adopting $\rho=20$. We show separately the contributions from the low and high extragalactic components (long and short dashes)  as well as the total one (solid lines). Two different low-energy extrapolations are displayed, with $\alpha_0=3.4$ and 2, and adopting the scaling of the nuclear cross sections with $a=1$ (red lines) or $a=2/3$ (blue lines). Also shown are the bounds from Figure~\ref{fwb} for the SFR case assuming all CRs are protons (purple line) and the WB bound (horizontal dashed line). Measurements from IceCube \cite{ic} are also shown. } 
\label{fig2}
\end{figure}

The results are shown in Figure~\ref{fig2}, where we compare the fluxes obtained in the previous section for the SFR scenarios under the assumption that the extragalactic sources emit just protons, with the ones obtained in the scenario from ref.~\cite{mo20} which involves two extragalactic populations with a mixed CR composition. For definiteness we adopt the model in which the component dominating below few EeV (referred to as `low') has a source evolution following the SFR, while the one dominating at higher energies (referred to as `high') has no evolution, since in this case the agreement with the observations is slightly  better. In this  SFR-NE scenario, the low extragalactic component has a source
 spectral index $\alpha=3.4$, with the relative fractions at the sources of the representative elements H, He, N, Si and Fe being $f_{\rm H}=0.19$, $f_{\rm He}=0.51$, $f_{\rm N}=0.30$ and $f_{\rm Si}=f_{\rm Fe}=0$, with a loosely determined rigidity cutoff, taken as 100\,EV. The high component dominating at the highest energies has instead $\alpha=2$ and $f_{\rm H}=0$, $f_{\rm He}=0.52$, $f_{\rm N}=0.30$,  $f_{\rm Si}=0.07$ and $f_{\rm Fe}=0.11$, with a relatively small rigidity cutoff at 1.4\,EV (see \cite{mo20} for more details). In the absence of magnetic horizon effects, the high component would have contributed to the CR flux at 1\,EeV twice as much as the low component, but due to the magnetic suppression its contribution to the observed CR flux becomes already subdominant below about 2\,EeV. We show in the Figure~\ref{fig2} the astrophysical neutrino flux upper bounds for the two cases corresponding to the nuclear cross sections scaling with $a=1$ and $a=2/3$, adopting in all cases $\rho=20$. We also show the two possible extrapolations of the flux of the low component that we discussed above. One can see that the sensitivity to the parameter $a$ is not large, affecting by a factor of about two the contribution from the high-energy component, which is mostly heavy, while the contribution from the low component, which consists mostly of light nuclei, is only slightly affected. 
 
 For comparison, we also include the measurements from the IceCube observatory, which below few PeV lie not far from the WB upper bound for SFR evolution (dashed horizontal line). Note that the flux that would be expected from thin sources, being proportional to $F_\pi\simeq \kappa P\leq P/2$, with $P<1$, should be  at least a factor of two below the horizontal  line that assumes $F_\pi=1$, in some tension with the observations. In this context the observations would suggest that the neutrino sources act, at least below 100~PeV, as CR reservoirs, magnetically confining the CRs below those energies for long enough times so that they lose most of their energies to pion production (see \cite{ah19} for a discussion). If one instead considers the more realistic scenario in which one of the extragalactic components extends with a steep spectrum well below the ankle energy, the 
 upper bound on the neutrino fluxes is well above the observations by IceCube. 
 This implies that in these scenarios the
 associated astrophysical neutrino  flux would be compatible with sources which are still thin to the escaping CRs. 
One should note that the relation between the upper bounds shown in Figure~\ref{fig2} and the actual neutrino fluxes would just reflect the (energy dependent) average column densities of the target material that would be traversed by the escaping CRs at the sources, or eventually in their environments. It would also reflect the energy dependence of the interactions, which is  strong for the $p\gamma$ scenarios for which one expects in particular a significant suppression below the threshold for $\Delta$ production. 
Let us  mention that a suppressed neutrino production will also have associated a suppressed photon production  through the decays of $\pi^0$ at the source, which could then be in agreement with bounds on the  GeV--TeV photons produced through  electromagnetic cascades of PeV photons with CMB background photons. Anyway it is relevant to also mention that the requirement of not   overpredicting the flux  of photons in the GeV--TeV range could constrain the steepness of the neutrino spectrum below 1~PeV \cite{mal}, and hence in $pp$ scenarios, in which the neutrino and photon source spectra tend to follow the shape of the CR spectrum, this could also constrain the steepness of the CR spectrum below 50~PeV. 

We note that $F_\pi$ is the fraction of the CR proton energies, and eventually also approximately that of the nucleons inside the nuclei, lost to pion production. Nuclei get also spallated or photodisintegrated as a consequence of the interactions, and the optical depth for these interactions is larger given that it is enough that one or few nucleons interact for these reactions to occur.
Clearly the possibility of having sources that are  thin to escaping protons could also allow for the heavy CRs to exit the sources without suffering excessive disintegrations, which is probably a desirable feature. Indeed,  while there is no evidence requiring the presence of nuclei from the Si or Fe group elements in the extragalactic fluxes below the ankle, the presence of  some CNO and He seems to be necessary.
The requirement that the sources be thin to pion production could be particularly relevant for $p\gamma$ scenarios, in which the abundant low energy photons could enhance the photo-disintegrations through the giant-resonance process.
Allowing for the sources to be thin gives more freedom to the scenarios that could explain the level of the observed astrophysical neutrino fluxes. In particular, it should allow to have a distribution of sources with different optical depths to pion production, so that not all have to be close  to the maximum allowed flux, and would not  require the need to eventually consider  contributions from hidden sources that produce neutrinos but not cosmic rays.

\section*{Acknowledgments}
This work was supported by CONICET (PIP 2015-0369) and ANPCyT (PICT 2016-0660). 
I am grateful to S.~Mollerach and F.~Vissani for comments.


\end{document}